\documentstyle[12pt]{article}
\begin{document}
\newcommand{\bea}{\begin{eqnarray}}
\newcommand{\eea}{\end{eqnarray}}
\newcommand{\be}{\begin{equation}}
\newcommand{\ee}{\end{equation}}
\newcommand{\non}{\nonumber}
\global\parskip 6pt
\begin{titlepage}
\vskip .5in
\begin{center}
{\Large\bf Subdivision Invariant Models in Lattice }\\
\vskip .25in
{\Large\bf Gauge Theory}\\
\vskip 1in
Danny Birmingham \footnote{Email: Dannyb@phys.uva.nl}     \\
\vskip .10in
{\em Universiteit van Amsterdam, Instituut voor Theoretische Fysica,\\
1018 XE Amsterdam, The Netherlands} \\
\vskip .50in
Mark Rakowski\footnote{Email: Rakowski@yalph2.bitnet}   \\
\vskip .10in
{\em Yale University, Center for Theoretical Physics,\\ New Haven,
CT 06511, USA}  \\
\end{center}
\vskip .10in
\begin{abstract}
   A class of lattice gauge theories is presented which exhibits novel
   topological properties. The construction is in terms of compact Wilson
   variables defined on a simplicial complex which models a four dimensional
   manifold with boundary.
   The case of $Z_{2}$ and $Z_{3}$ gauge groups is considered in detail, and
   we prove that at certain discrete values of the coupling parameter,
   the partition function in these models remains
   invariant under subdivision of the underlying
   simplicial complex.
   A variety of extensions is also presented.

\end{abstract}
\vskip .5in
\begin{center}
ITFA-93-03\,/\, YCTP-P4-93 \\
February 1993
\end{center}
\end{titlepage}

\section{Introduction}
 In this paper, we undertake a study of certain lattice gauge theories
which have special properties with respect to subdivision of the
underlying lattice. The motivation for such a search has
its roots in topological field theory (see \cite{BBRT} for a review),
where quantum field theories have been constructed whose observables
are topological or smooth invariants of the underlying spacetime manifold.
In the pure Chern-Simons theory \cite{AS,EWit}, one
has, in particular, a partition function which is a topological invariant
of a framing of the spacetime 3-manifold. Our interest originates primarily
from the desire to see these types of structures emerge from a traditional
lattice approach.

  Calculations in lattice gauge theory, and statistical mechanics generally,
are concerned with the behaviour of systems in a continuum limit, where the
underlying lattice  is subdivided into smaller and smaller units. At any given
stage of subdivision, one has only  a crude approximation to
the continuum theory.
Topological field theories are, on the other hand, quite
different. The topology of any manifold can be captured in terms of a
lattice (simplicial complex), and further subdivisions of that lattice in
no way enhance ones topological picture of the space. It is of interest to
construct lattice models which also reflect this property; models in which
the observables are invariant under lattice subdivision. There
is then no need to be concerned with a continuum limit, as the model would
already compute - exactly - the relevant quantities. In other words, one
would already be at the continuum limit.

   Here, we construct models which have the property of subdivision
invariance at certain descrete values of the coupling parameter. While
our motivation for these particular examples stems primarily from the
Chern-Simons theory, we will not establish any firm link with that theory
here. Our approach is entirely self-contained and we will have no need to
refer to results in any continuum model, or to invoke general folklore
in quantum field theory. Although nothing
in our construction forces us to consider discrete gauge groups,
our analysis here
will focus on these simpler examples. We find that the partition function
of our model, which is defined on a simplical complex which models a 4-manifold
with boundary, is invariant under the type 4 Alexander subdivision \cite{Alex}.
This
is essentially a local property which we can prove by looking at the
Boltzmann weight on a single 4-simplex. This special property is
restricted to discrete values of the coupling parameter. We also consider other
types of lattice subdivision, and show that the
partition function of our theory on a disk is invariant under all of the
Alexander moves.

We begin in the next section with an overview of lattice gauge theory in
terms of compact Wilson variables, and provide some background on simplicial
complexes. Our model is then defined and we move on to consider specific
cases in succeeding sections. A simple two dimensional version is considered
first, and then we treat the $Z_{3}$ and $Z_{2}$ gauge groups in four
dimensions. While all our detailed calculations are for discrete groups,
we discuss some obvious extensions both to continuous groups, and to higher
dimensional analogs of the models presented here.
We close then with some concluding remarks.

\section{General Properties}
We first recall the essential definitions
needed in a Wilson formulation of lattice gauge theory on a simplicial
complex. For a complete account of the latter, see \cite{JM}.

Let $[v_{0},\cdots ,
v_{n}]$ denote the oriented n-simplex spanned by the geometrically
independent set of points $\{ v_{i} \}$, called its vertices. One can
picture these simplices as points, line segments, triangles and
tetrahedrons for n equal to zero through three. A simplex which is
spanned by any subset of the vertices is called a face of the original
simplex.

A simplicial complex $K$
is a collection of simplices which are glued together under two
restrictions. Any face of a simplex in $K$ is required to be a simplex
in $K$, and the intersection of any two simplices in $K$ must be a face
of each of them.

The basic fields which enter a formulation of lattice
gauge theory  are group valued maps on the 1-simplices (denoted $[a,b]$)
with the rule
that $U_{ba} = U^{-1}_{ab}$.
In order to define a theory, one requires that the group be compact
with an invariant measure, and one takes the action on the ``link''
variables - the gauge transformations - to be given by:
\bea
U_{ab} \rightarrow g_{a}\; U_{ab}\; g^{-1}_{b}\;\;.
\eea
Here, $g_{a}$ is a group element associated with the vertex $a$.

We take the action of our theory to be a gauge invariant function $S$ of
the above link variables. The partition function is then
\bea
Z = \prod_{\alpha} \int \; dU_{\alpha}
\;\exp [\beta\, S(U) ]\;\;,\label{part}
\eea
where the index $\alpha$ indicates the set of independent $1$-simplices.
In the case of a discrete gauge group, the group integration (whose volume
we normalize to unity) is a discrete sum,
\bea
\int \, dU \rightarrow \frac{1}{|G|}\, \sum_{U}\;\; ,
\eea
where $|G|$ denotes the order of the group.
One can further define correlation functions of the link variables,
\bea
<U_{\gamma_{1}}\cdots U_{\gamma_{p}} > = \prod_{\alpha} \int\; dU_{\alpha}\;
U_{\gamma_{1}}\cdots U_{\gamma_{p}}\; \exp [ \beta\, S(U) ]\;\;.
\eea
In the above, a group trace is understood, whenever necessary.
The key point is that, in general,
all of these quantities depend not only on the
coupling parameter $\beta$, but also on the simplicial complex $K$.

The above was an abstract description of lattice gauge theory; however,
the motivation
for it arises as follows. Normally in a gauge theory the basic
dynamical variable is a connection on some principal $G$ bundle over
spacetime. One can construct the Wilson link variables
\bea
W_{C_{ab}} = P\;\exp [ \int_{C_{ab}}  A ]\;\;,\label{WL}
\eea
where $C_{ab}$ is some path in spacetime between the endpoints $a$ and $b$,
and $A$ is a connection which takes
values in the Lie algebra of the group $G$. As usual, the symbol $P$
denotes the path ordering.
It is straightforward to establish that the the link variables are
solutions to the differential equation:
\bea
D\; W_{C_{ab}} = 0\;\;,
\eea
where $D$ is the covariant derivative.

Let $U_{abc} = U_{ab}\, U_{bc} \, U_{ca}$ be the holonomy,
based at the first
vertex $a$, around  the triangle determined by $a$, $b$ and $c$, and
traversed in the order from left to right .
In terms of the Wilson link variables, this is represented by
choosing a closed contour $C$ in (\ref{WL}), and taking a group
trace if necessary.
A property of holonomy is that in the limit when the loop becomes
infinitesimally small, $U_{C}$ approaches
\bea
\exp [ \int  F ]\;\;,
\eea
where the integral is over a surface which has $C$ as its boundary.

The usual Wilson action is given by
\be
S = \frac{1}{2}\;\sum_{U}\;(U - 1) + (U^{-1} -1)\;\;,
\ee
where the sum is over all the basic holonomies on the lattice.
Notice that each term in this sum depends on a single holonomy.
It is a standard exercise \cite{Creutz} to show that the continuum limit
of the above action is the Yang-Mills theory:
\be
S \rightarrow \frac{1}{2}\int\; F^{2}\;\;.
\ee

In the present discussion, we are motivated to consider a discrete
version of the Chern form, $F\wedge F$. The first observation
is to note that one can produce $F$
by combinations such as
$U-U^{-1}$ and $U-1$, in the continuum limit. We will base our action
on a combination of two independent holonomies,
which are tied together at a point,
and have no edges in common. Let us first, however, digress to review
a product which will serve to form the analog of the wedge product of
differential forms.

Denote
by $P$, one of the $(r+s+1)!$ permutations of the set of vertices $\{
v_{0},\cdots,v_{r+s}\}$, which span some $(r+s)$-simplex,
and by $Pv_{i}$ the value of that permutation on
$v_{i}$.
Let $c^{r}$ and $c^{s}$ be group valued maps on $r$- and $s$-simplices,
respectively. The $\star$-product $c^{r}\star c^{s}$ yields a group
valued map defined on $(r+s)$-simplices, and is defined by:
\pagebreak
\bea
&\phantom{.}& < c^{r}\star c^{s}, [v_{0},\cdots,v_{r+s}] > \;
= \label{star}\\
&\phantom{.}& \frac{1}{(r+s+1)!}\,
\sum_{P}\; (-1)^{| P |}\; < c^{r},[Pv_{0},\cdots,Pv_{r}] >\;
\cdot < c^{s},[Pv_{r},\cdots, Pv_{r+s}] >\;\;,\nonumber
\eea
when the order $v_{0}\cdots v_{r+s}$ is in the equivalence class of the
orientation of the simplex $[v_{0},\cdots,v_{r+s}]$ (this determines the
overall sign of the product), and
where the sum is over all permutations of the vertices.
The notation $\langle \cdot \, , \cdot \rangle$ is used to indicate the
evaluation of the map on the accompanying simplex, and the product
on the right hand side of (\ref{star}) refers to multiplication in
the relevant group or ring.
For a more complete definition of the $\star$-product, we refer
to \cite{BR}; we simply note here that it is a variation of the standard
cup product which achieves graded commutativity at the expense of
associativity (the usual cup product is graded commutative only on
cohomology classes).

The actual number
of independent terms in the above  sum is given
by the number of ways one can
partition the set of vertices into two parts which
contain one vertex in
common, and an easy counting yields
\bea
\frac{(r+s+1)!}{r!\, s!}\;\;.
\eea

Now we are equipped to return to our action.
We take this to be a sum over holonomy pairs
\bea
S= \sum \; (U - U^{-1})\star (U - U^{-1})\;\; ; \label{a1}
\eea
a trace is understood when required, and the
sum here is over all elementary $4$-simplices in the simplicial
complex.
The $\star$-product
ensures that the two factors are independent holonomies
with one point in common. Of course,
one could equally well consider an action of the form
\bea
S = \sum\; (U-1)\star (U-1)\;\;,                      \label{a2}
\eea
or several other possibilities, each of which
yield the same
continuum limit. Note, however, that the quantity $(U - U^{-1})_{abc}$
has the additional feature that it is antisymmetric in its last two indices;
this follows simply from $U_{acb} = U^{-1}_{abc}$.
As we have seen, the holonomy $U$ is a group valued map on $2$-simplices.
The actions defined above, in terms of two independent holonomies,
are therefore naturally defined on a $4$-simplex.

We should emphasize that achieving a continuum limit akin to the form
$F \wedge F$ is purely motivational, and we shall not draw on any
properties of continuum theories here.
It is not clear, a priori,
whether any of the familiar continuum properties will translate
onto a finite lattice.
We note that an action similar to the above was studied for the case
of a torus, in a different context \cite{Ven}.
Our aim is simply to take the above lattice
definition, and prove that it has certain topological features.

When we evaluate the star product on a given 4-simplex, the
antisymmetrization produces generically 5! terms. With the action
(\ref{a1}), the symmetries present reduce that to 15 distinct terms.
We will take as the Boltzmann weight of this theory, evaluated on
the simplex $[v_{0},v_{1},v_{2},v_{3},v_{4}]$, the normalization
given by:
\bea
W[0,1,2,3,4]&=& B[0,1,2,3,4]\,B[0,1,3,4,2]\,B[0,1,4,2,3]\non\\
& &B[1,0,2,4,3]\,B[1,0,3,2,4]\,B[1,0,4,3,2]\non\\
& &B[2,0,1,3,4]\,B[2,0,3,4,1]\,B[2,0,4,1,3]\non\\
& &B[3,0,1,4,2]\,B[3,0,2,1,4]\,B[3,0,4,2,1]\non\\
& &B[4,0,1,2,3]\,B[4,0,2,3,1]\,B[4,0,3,1,2]\;\;,\label{bw}
\eea
where,
\bea
B[0,1,2,3,4] = \exp [\beta\, (U-U^{-1})_{v_{0}v_{1}v_{2}}\,
(U-U^{-1})_{v_{0}v_{3}v_{4}} ]\;\; .
\eea
Our analysis deals with the general case of complex coupling $\beta$.
We will, however, still use the phrase ``Boltzmann weight" in
this more general context.

Using the Boltzmann weight defined above, we can compute the partition
function for the theory defined on  a simplicial complex $K$.
The central issue of interest here is to examine how this function behaves
upon subdivision of the complex. For ease of illustration, let us consider
a single $4$-simplex $[v_{0},v_{1},v_{2},v_{3},v_{4}]$.
A convenient basis of
subdivision operations,
known as Alexander moves \cite{Alex}, are available, and these allow a direct
analysis of this question.
The  Alexander
moves can be described in turn by:

Type 1 Alexander subdivision:
\begin{eqnarray}
[ v_{0},v_{1},v_{2},v_{3},v_{4} ] \rightarrow [ x,v_{1},v_{2},v_{3},
v_{4} ] + [v_{0},x,v_{2},v_{3},v_{4} ]\;\;,
\end{eqnarray}

Type 2 Alexander subdivision:
\begin{eqnarray}
[ v_{0},v_{1},v_{2},v_{3},v_{4}] \rightarrow
[ x,v_{1},v_{2},v_{3},v_{4} ] + [ v_{0},x,v_{2},v_{3},
v_{4} ] + [v_{0},v_{1},x,v_{3},v_{4} ]\;\;,
\label{as2}
\end{eqnarray}

Type 3 Alexander subdivision:
\begin{eqnarray}
[ v_{0},v_{1},v_{2},v_{3},v_{4}] &\rightarrow&
[ x,v_{1},v_{2},v_{3},v_{4} ] + [ v_{0},x,v_{2},v_{3},
v_{4} ] + [v_{0},v_{1},x,v_{3},v_{4} ]\non\\
&+& [ v_{0},v_{1},v_{2},x,v_{4} ]\;\;,
\end{eqnarray}

Type 4 Alexander subdivision:
\begin{eqnarray}
[ v_{0},v_{1},v_{2},v_{3},v_{4}] &\rightarrow&
[ x,v_{1},v_{2},v_{3},v_{4} ] + [ v_{0},x,v_{2},v_{3},
v_{4} ] + [v_{0},v_{1},x,v_{3},v_{4} ] \non\\
&+& [ v_{0},v_{1},v_{2},x,v_{4} ] + [ v_{0},v_{1},v_{2},v_{3},x ]\;\;.
\end{eqnarray}

One can picture the move of type $1$ as the introduction of an
additional vertex
$x$, which is placed at the center of the $1$-simplex
$[v_{0},v_{1}]$, and is then joined to all the remaining vertices
of the $4$-simplex.
Moves $2$ to $4$ involve a similar construction, where $x$ is placed
at the center of the simplices $[v_{0},v_{1},v_{2}]$, $[v_{0},v_{1},v_{2},
v_{3}]$, and finally $[v_{0},v_{1},v_{2},v_{3},v_{4}]$.
There is, in addition, a type $0$ move which is effected by replacing
a vertex of the simplicial complex by a new vertex. This can be
considered as  a degenerate case, and need not concern us in the following.

According to Alexander \cite{Alex},
two simplicial complexes are said to be equivalent if and only if it is
possible to transform one into the other by a sequence of these moves.
Hence, any function of $K$ which is invariant under these moves
yields a combinatorical invariant of the simplical complex.

\section{A Toy Model in Two Dimensions}

   Let us consider a simpler two dimensional version of the type of models
we want to explore. Here, it is natural to consider an action which
depends on a single holonomy, and for the gauge group
$Z_{3}=\{1,\exp[2\pi i/3], \exp[4\pi i/3]\}$,
we will
take the Boltzmann weight evaluated on the 2-simplex $[v_{0},v_{1},v_{2}]$
to be given by:
\bea
W[v_{0},v_{1},v_{2}] =
\exp [ \beta\, (U - U^{-1})_{v_{0}v_{1}v_{2}} ] \;\; ,\label{bw2d}
\eea
where  $U_{v_{0}v_{1}v_{2}}$ is the holonomy combination $U_{v_{0}v_{1}}
U_{v_{1}v_{2}} U_{v_{2}v_{0}}$.

   It is now a simple matter to explore the subdivision properties of this
theory. First, consider the Boltzmann weight under the type 2 Alexander
move where
\bea
[v_{0},v_{1},v_{2}] \rightarrow [v_{3},v_{1},v_{2}] + [v_{0},v_{3},v_{2}]
+ [v_{0},v_{1},v_{3}] \;\; .
\eea
The picture of this subdivision operation is simply that of adding a new
$v_{3}$ vertex to the center of the original, and then connecting that
to the other vertices by the addition of three new links. This present theory
enjoys the special property that:
\bea
W[v_{0},v_{1},v_{2}] = \frac{1}{27}\;\sum_{U_{v_{0}v_{3}},U_{v_{1}v_{3}},
U_{v_{2}v_{3}}}\; W[v_{3},v_{1},v_{2}]\; W[v_{0},v_{3},v_{2}]\;
W[v_{0},v_{1},v_{3}] \;\;,
   \label{a2d2}
\eea
when the coupling takes values such that $s^{3} = 1$,
where $s = e^{i\,\beta \sqrt{3}}$
is a convenient scale parameter. This property is not obvious, but can be
checked, by hand, in a straightforward way. Notice that at the special
points, the Boltzmann weight itself is $Z_{3}$-valued for every link
configuration, and this will be the pattern in all our examples.

  The situation under type 1 Alexander subdivision is slightly trickier.
Here, a given 2-simplex $[v_{0},v_{1},v_{2}]$ is broken into two pieces;
\bea
[v_{3},v_{1},v_{2}] + [v_{0},v_{3},v_{2}] \;\; .
\eea
The picture is that of splitting the $[v_{0},v_{1}]$ edge by the introduction
of the $v_{3}$ vertex at its center, and then connecting that to $v_{2}$
with a new link. One can ask how the Boltzmann weight or partition
function behaves under this move. Again, one can verify by hand that
\bea
W[v_{0},v_{1},v_{2}] = \frac{1}{3} \; \sum_{U_{v_{2}v_{3}}} \;
W[v_{3},v_{1},v_{2}]\; W[v_{0},v_{3},v_{2}]\;\;,
 \label{a1d2}
\eea
when $s^{3} = 1$ and $U_{v_{0}v_{1}} = U_{v_{0}v_{3}} U_{v_{3}v_{1}}$.
We were fortunate in this case to have a simple relation between the
boundaries of the original simplex, and that of the simplices after
type 1 subdivision; the extra constraint on the product of two
of the new link variables just reflects that relationship.

Of course, the action (\ref{bw2d}) vanishes for the group $Z_{2}$, and so,
in line with the analysis of the previous section, one may wish
to examine the $Z_{2}$ model with Boltzmann weight:
\be
W[v_{0},v_{1},v_{2}] = \exp [ \beta\, (U - 1)_{v_{0}v_{1}v_{2}} ] \;\; .
\ee

The immediate observation, however, is that the above is simply the standard
Wilson action for this group. Nevertheless, if one examines this model
with a complex coupling $\beta$, then a simple analysis reveals
that the Boltzmann weight is invariant under the Alexander moves of type $1$
and $2$. This takes places when the coupling satisfies
$s^{2}=1$, where $s=e^{-2\beta}$, and
$U_{v_{0}v_{1}}=U_{v_{0}v_{3}}U_{v_{3}v_{1}}$.

It is well known that two-dimensional Yang-Mills
theory is invariant under area-preserving subdivisions, although such a result
is obtained for $\beta$ real and positive, and using a heat kernel form
of the action \cite{Mig,EWit2}.
This simple example indicates that it might be fruitful to re-examine
the usual Wilson action and search for special subdivision properties
at complex couplings.

One other point that is especially transparent in this model concerns
our choice to include the $-1$ term in the defining action; one might
ask whether that really has any significance. Suppose we had defined
the Boltzmann weight
\bea
W'[v_{0},v_{1},v_{2}] = \exp [ \beta\, U_{v_{0}v_{1}v_{2}} ]\;\; .
\eea
Then a simple scaling of our previous result yields the
relation,
\bea
W'[v_{0},v_{1},v_{2}] = \frac{1}{8}\;s \sum_{U_{v_{0}v_{3}},U_{v_{1}v_{3}},
U_{v_{2}v_{3}}}\; W'[v_{3},v_{1},v_{2}]\; W'[v_{0},v_{3},v_{2}]\;
W'[v_{0},v_{1},v_{3}]\;\;,
\label{s}
\eea
which we can interpret as subdivision invariance up to a scaling factor.
This lends some support to our original geometrical motivation which
suggested the combination $(U - 1)$. If one were to search more generally
for other models with interesting subdivision properties, it would be
important not to discard models which had this additional scaling behavior.

\section{The $Z_{3}$ Model}

We will consider in this section the type of model we outlined
in the section on General Properties; i.e., a gauge theory whose action
is based on two independent holonomies which are tied together at a
point and have no edges in common. Pictorially, one might refer to
this as a ``bowtie" configuration.
The setting here is on a simplicial
complex which models a four dimensional manifold with boundary (which
can also be empty) and we will focus on the group $Z_{3}$, which we
represent multiplicatively as the cube roots of unity, $1$,
$\exp [2\pi i /3]$ and $\exp [-2\pi i /3]$. We were motivated in our choice
of action by the familiar Chern form, and we saw earlier that we had
some freedom in writing a discrete analog. Our calculations in this
section will be based on the action (\ref{a1}), and we will take the
Boltzmann weight for a given ordering of vertices to be given by:
\bea
B[0,1,2,3,4]=\exp [\beta (U - U^{-1})_{v_{0}v_{1}v_{2}}
\; (U-U^{-1})_{v_{0}v_{3}v_{4}} ]\;\; ,
\eea
and we will insert that into the expression (\ref{bw}) to get a quantity
$W[0,1,2,3,4]$ which takes into account all the different permutations
of the  $\star$-product. It will be useful in the following analysis
to introduce a scale parameter which we take in this model to be
the quantity $ s = \exp[-3 \, \beta] $.

The behaviour of the theory under the type 4 Alexander move parallels
that of the simple two dimensional model under the type 2 move. Let
us investigate the Boltzmann weight of this model in the same way.
Take the 4-simplex $[v_{0},v_{1},v_{2},v_{3},v_{4}]$ and its
corresponding type 4 subdivision which is the sum of five 4-simplices:
\bea
& &[v_{5},v_{1},v_{2},v_{3},v_{4}] + [v_{0},v_{5},v_{2},v_{3},v_{4}] +
[v_{0},v_{1},v_{5},v_{3},v_{4}] + \non \\
& &[v_{0},v_{1},v_{2},v_{5},v_{4}] + [v_{0},v_{1},v_{2},v_{3},v_{5}]
\;\; .\label{a4s}
\eea
The property of this theory is that
\bea
W[v_{0},v_{1},v_{2},v_{3},v_{4}] = \frac{1}{3^{5}}\; \sum_{U_{i5}}
W[v_{5},v_{1},v_{2},v_{3},v_{4}]\;
W[v_{0},v_{5},v_{2},v_{3},v_{4}]& &\non\\
W[v_{0},v_{1},v_{5},v_{3},v_{4}]
W[v_{0},v_{1},v_{2},v_{5},v_{4}]
W[v_{0},v_{1},v_{2},v_{3},v_{5}]\;\;,& &
\label{b4s}
\eea
when $s^{3}=1$ and
where the sum is over the 5 links which join to the new $v_{5}$ vertex.
We do not know an elementary way of seeing this at the present time. However,
it is simple enough to write a computer program to check this kind of
relation and verify it for all choices of ``boundary data", and this we have
done using Mathematica \cite{Math}. The relation is quite simple owing
to the fact that both the boundary of the original simplex and its type
4 subdivision are identical. This property guarantees that when we
evaluate the partition function of the theory on any simplicial complex,
that the number one gets is invariant under all type 4 subdivisions.
We emphasize again that this special property only holds for the
values of the coupling $\beta$ such that $s^{3}=1$.

While we do not yet have such a complete understanding
of the subdivision properties of this model under the other Alexander moves
at the level of Boltzmann weights, we can nevertheless offer some
computational evidence why it is interesting. Here, we will compute
exactly the partition function of the theory on the 4-disk which provides
an example with a boundary topologically equivalent to $S^{3}$. One
can model the disk as a simplicial complex with a single 4-simplex, or
through more complex subdivisions.

The following results for the
partition function (\ref{part}) were calculated using Mathematica. One
aspect of lattice gauge theory that  is important to take advantage
of in these computer studies is the freedom to gauge fix some link
components. The issue of gauge fixing in the Wilson
formulation is particularly
simple and elegant and does not introduce any murky questions which
could undermine the rigour of our analysis. The construction of the
partition function in terms of group integrations and a gauge invariant
action allows one to fix arbitrarily the links on a maximal tree
\cite{Creutz}. Roughly speaking, this is any collection of
links which does not include a closed path and which cannot be extended
by the addition of other links. From a practical perspective, this
significantly reduces the number of group integrations (which are just
finite sums in this case) that we must perform.  Let us now list, in
turn, the results of our calculation for the 4-disk.

For the representation of the disk in terms of a single 4-simplex,
we find the partition function:
\begin{eqnarray}
Z = \frac{1}{3^{6}}\, (221 + 120 (s+s^{-1}) + 60 (s^{2} + s^{-2}) +
    54 (s^{5}+s^{-5}) + 20 (s^{6}+s^{-6}))\;\; .\non\\
\label{z31}
\end{eqnarray}
Under a type 1 Alexander subdivision of that simplex, we find:
\begin{eqnarray}
Z&=&\frac{1}{3^{9}}\, (4215 + 2256 (s+s^{-1}) + 1596 (s^{2}+s^{-2}) +
  720 (s^{3}+s^{-3})\non\\
&+& 660 (s^{4}+s^{-4}) + 1068 (s^{5}+s^{-5}) +
  664 (s^{6}+s^{-6}) + 456 (s^{7}+s^{-7})\non\\
&+& 48 (s^{8}+s^{-8}) +
  162 (s^{10}+s^{-10})
+ 72 (s^{11}+s^{-11}) + 32 (s^{12}+s^{-12}))\;\;.\non\\
\label{z32}
\end{eqnarray}
Under a type 2 Alexander subdivision, the computation yields:
\begin{eqnarray}
Z &=& \frac{1}{3^{10}}\,(9641 + 5544 (s+s^{-1}) + 4482 (s^{2}+s^{-2}) +
   2610 (s^{3}+s^{-3})\non\\
&+& 2178 (s^{4}+s^{-4}) + 3222 (s^{5}+s^{-5}) +
   2286 (s^{6}+s^{-6}) + 1764 (s^{7}+s^{-7})\non\\
&+& 738 (s^{8}+s^{-8}) + 522 (s^{9}+s^{-9})
+ 666 (s^{10}+s^{-10}) + 270 (s^{11} + s^{-11})\non\\
&+& 294 (s^{12}+s^{-12}) + 54 (s^{13}+s^{-13}) + 30 (s^{15}+s^{-15})
+ 18 (s^{16}+s^{-16})\non\\
&+& 18 (s^{17}+s^{-17}) + 8 (s^{18}+s^{-18}))\;\; .
\label{z33}
\end{eqnarray}
For the case of the disk represented by four 4-simplices which are
the type 3 subdivision of the original simplex, we have:
\bea
Z &=& \frac{1}{3^{10}}\,(10293 + 4680 (s+s^{-1}) + 3756 (s^{2}+s^{-2}) +
2064 (s^{3}+s^{-3})\non \\
&+& 2508 (s^{4}+s^{-4}) + 2544 (s^{5}+s^{-5}) + 1840 (s^{6}+s^{-6}) +
1992 (s^{7}+s^{-7}) \non\\
&+& 1638 (s^{8}+s^{-8}) + 1104 (s^{9}+s^{-9}) + 1080 (s^{10}+s^{-10}) +
600 (s^{11}+s^{-11}) \non\\
&+& 320 (s^{12}+s^{-12}) + 72 (s^{13}+s^{-13}) + 60 (s^{14}+s^{-14}) +
16 (s^{15}+s^{-15}) \non\\
&+& 24 (s^{16}+s^{-16}) + 72 (s^{18}+s^{-18}) + 8(s^{21}+s^{-21}))\;\; ,
\eea
and finally for the partition function on the simplical complex resulting
from the fourth Alexander move, we have:
\bea
Z&=&\frac{1}{3^{10}}\; (11841 + 5460 (s+s^{-1}) + 2640 (s^{2}+s^{-2}) +
780 (s^{3}+s^{-3}) \non\\
&+& 2250 (s^{4}+s^{-4}) + 2034(s^{5}+s^{-5}) +600(s^{6}+s^{-6}) +
1560(s^{7}+s^{-7}) \non\\
&+& 2520 (s^{8}+s^{-8}) + 2970 (s^{9}+s^{-9}) + 1560 (s^{10}+s^{-10}) +
600 (s^{11}+s^{-11})\non\\
&+& 180(s^{12}+s^{-12}) + 90 (s^{13}+s^{-13}) + 60 (s^{15}+s^{-15}) +
60 (s^{16}+s^{-16}) \non\\
&+& 120 (s^{17}+s^{-17}) + 60(s^{18}+s^{-18}) + 60(s^{19}+s^{-19}))\;\; .
\label{zz3alex4}
\eea

Although these results may appear at first glance to have neither
rhyme nor reason, we expect that they will have special properties at
the points $s^{3}=1$. Indeed, all five of the functions in
(\ref{z31}) - (\ref{zz3alex4}) reduce to the simple form,
\begin{eqnarray}
Z(s) =\frac{1}{3^{4}}\; (\, 29 + 26\, (s + s^{-1}) )\;\; ,
\end{eqnarray}
at those particular values of $s$. When $s=1$, or equivalently $\beta = 0$,
we have a trivial subdivision invariant point, and $Z=1$, but at the
other two cube roots of unity, we find the value $Z= 1/27$.

As a second example, let us consider the $4$-dimensional sphere $S^{4}$.
We model $S^{4}$ as the boundary of a 5-simplex:
\bea
\partial [v_{0},v_{1},v_{2},v_{3},v_{4},v_{5}] &=& [v_{1},v_{2},v_{3},
v_{4},v_{5}] + [v_{0},v_{3},v_{2},v_{4},v_{5}] + [v_{0},v_{1},v_{3},
v_{4},v_{5}]\non\\
&+& [v_{0},v_{1},v_{2},v_{5},v_{4}] + [v_{0},v_{1},v_{2},
v_{3},v_{5}] + [v_{0},v_{1},v_{2},v_{4},v_{3}]\;\;,\non\\
\label{s4}
\eea
where we have chosen to write it in a way such that each 4-simplex is
positiviely oriented according to the order given by its vertices.
The precise definition of the boundary operator $\partial$ can be found
in \cite{JM}.
The result of this computation is
\be
Z=\frac{1}{3^{10}}
(33309 + 12300(s^{9} +s^{-9}) + 570(s^{18} +s^{-18}))\;\; .
\ee
When we restrict $s$ to be a cube root of unity, we find that $Z=1$.

\section{A Novel $Z_{2}$ Model}

We already considered a two dimensional $Z_{2}$ based model in an
earlier section; here we would like to extend that to four dimensions.
Of course, actions which depend on the combination $U-U^{-1}$ necessarily
lead to a trivial theory for this group, but there is another problem
if we use the $U-1$ combination in concert with the $\star$-operator.
For abelian groups generally, the holonomy $U_{abc}$ is invariant
under cyclic permutations of the indices, so that the base point
of the holonomy does not enter. In the case of $Z_{2}$, we also
have that $U = U^{-1}$ for all group elements, so the holonomy
combination is in fact symmetric in all indices. We tacitly avoided
this in two dimensions and simply defined how to evaluate the Boltzmann
weight on a given 2-simplex. It is interesting that one can do something
similar in four dimensions as well, and we will take as a matter
of definition the expression (\ref{bw}) for $W[0,1,2,3,4]$ together
with a new quantity,
\bea
B[0,1,2,3,4] = \exp [ \beta\, (U-1)_{v_{0}v_{1}v_{2}}\, (U-1)_{v_{0}
v_{3}v_{4}} ]\;\; .
\eea
This has the effect of selecting essentially half the terms which
would appear in the $\star$-product, and ``discarding" those which
would have entered with opposite sign. The bottom line as to whether
this is a fruitful line of thought is whether we can achieve similar
success in terms of subdivision properties. As in all the models,
we will find a certain scale combination convenient, and we take
$s=\exp [ 4\,\beta ]$ in this section.

The first order of business is to analyze the subdivision properties
under the type 4 Alexander move.  It turns out that this model also enjoys
the simple relation (\ref{b4s}) for its Boltzmann weight at the
points $s^{2} = 1$. Although the number of link variables is much
smaller than in the $Z_{3}$ example, the number is
nevertheless quite large, and we also employed a computer
program to verify this claim.

It is also straightforward to analyze the other subdivision properties
of this model in the explicit calculation of the partition function
on a 4-disk. We find, in the same way, that for a single 4-simplex
this theory yields the partition function,
\begin{eqnarray}
Z = \frac{1}{2^{6}}\, (11 + 15 s^{2} + 27 s^{5} + 10 s^{6} + s^{15})\;\; .
\end{eqnarray}
Under the first Alexander move, where there are two 4-simplices, we
find,
\begin{eqnarray}
Z &=& \frac{1}{2^{9}}\,(31 + 54 s^{2} + 33 s^{4} + 48 s^{5} + 12 s^{6} +
    96 s^{7} + 24 s^{8} + 105 s^{10}\non\\
&+& 72 s^{11} + 22 s^{12} +
    6 s^{20} + 8 s^{21} + s^{30})\;\; .
\end{eqnarray}
The type 2 Alexander move applied to the original simplex leads to,
\begin{eqnarray}
Z &=& \frac{1}{2^{10}}\, (29 + 45 s^{2} + 54 s^{4} + 45 s^{5} + 27 s^{6} +
    108 s^{7} + 18 s^{8} + 72 s^{9} + 54 s^{10}\non\\
&+& 18 s^{11} + 153 s^{12} +
    36 s^{13} + 27 s^{14} + 102 s^{15} + 135 s^{16} + 45 s^{17} +
    13 s^{18}\non\\
&+& 9 s^{25} + 18 s^{26}
+ 12 s^{27} + 3 s^{36} + s^{45})   \;\; ,
\end{eqnarray}
while for the type 3 move we find,
\begin{eqnarray}
Z &=& \frac{1}{2^{10}}\, (19 + 12 s^{2}+ 48 s^{4} + 24 s^{5} + 16 s^{6} +
    48 s^{7} + 15 s^{8} + 72 s^{9} + 42 s^{10}\non\\
&+& 84 s^{12} + 48 s^{13} +
    96 s^{14} + 24 s^{15} + 30 s^{16} + 96 s^{17} + 28 s^{18} + 75 s^{20} +
    104 s^{21}\non\\
&+& 84 s^{22}
+ 4 s^{24} + 4 s^{30} + 24 s^{31} +
    12 s^{32} + 8 s^{33} + 6 s^{42} + s^{60})\;\; .
\end{eqnarray}
Lastly, the type 4 Alexander move, which we already know is an invariance
of the partition function at $s^{2}=1$ from our more general analysis,
leads to,
\begin{eqnarray}
Z &=& \frac{1}{2^{10}}\, (16 + 30 s^{4} + 15 s^{5} + 30 s^{6} + 60 s^{9} +
    45 s^{10} + 70 s^{12} + 90 s^{14} + 60 s^{15}\non\\
&+& 70 s^{18} +
    120 s^{19} + 30 s^{20} + 90 s^{22} + 27 s^{25} + 75 s^{26} +
    130 s^{27} + 20 s^{36}\non\\
&+& 30 s^{37} + 5 s^{39} +
    10 s^{48} + s^{75})\;\; .
\end{eqnarray}

All of these results become much more transparent when we restrict
them to the points where $s^{2}=1$. It is remarkable that all of the
above polynomials reduce to the simple formula,
\begin{eqnarray}
Z(s)=\frac{1}{2^{4}}\; (\, 9 + 7\, s )   \;\; .
\end{eqnarray}
The two roots of unity, $+1$ and $-1$ yield the values $1$ and $1/8$
respectively for the partition function of the disk.

Again, we can compute the  partition function on $S^{4}$, and in this instance
we find:
\bea
Z&=& \frac{1}{2^{10}}(
  16 + 60\,{s^6} + 45\,{s^{10}} + 15\,{s^{12}} + 180\,{s^{14}} +
   20\,{s^{18}} + 180\,{s^{20}} + 180\,{s^{24}}\non\\
&+& 45\,{s^{28}}
+ 27\,{s^{30}} + 180\,{s^{32}} + 60\,{s^{42}} + 15\,{s^{54}} + {s^{90})}\;\;.
\eea
Observe that the polynomial for $S^{4}$ contains even powers of $s$ only.
When $s^{2}=1$, the partition function therefore assumes the value $Z=1$.

\section{Generalizations}

Here, we discuss some generalizations and extensions of the models
considered above. The framework we have outlined is obviously quite
general, and one can immediately consider corresponding theories
based on the traditional continuous gauge groups.
The only constraint, as we have noted, is that the group should have
an invariant measure and finite volume, or new ideas are required.
On a finite lattice,
the partition function is a well defined number, and the issue is whether
these other gauge groups allow for special subdivision invariant points.
The behaviour of these theories in the continuum limit for generic
couplings is a far more
difficult question, and we will not address that here.
{}From a purely topological perspective, one might adopt the point of
view that behaviour away from the special points is irrelevant.

As our final example, let us consider a model in six dimensions.
Clearly, one can define a model in any even dimension, either with discrete
or continuous gauge groups.
The construction is quite straightforward, and simply involves taking
the star product of three independent holonomies, viz.,
\be
S =\sum \; (U-U^{-1})\star (U-U^{-1})\star (U-U^{-1})\;\;.
\ee
Such an action will be evaluated on a $6$-simplex,
and the sum is over all the elementary $6$-simplices
in the simplicial complex. Again, it would  be of
interest to examine  the subdivision properties of the
associated partition functions.

\section{Concluding Remarks}
As we have seen,
the interesting subdivision properties of the models presented here are
based upon a few key ingredients. In particular, one is guided to
employ the star product operator in seeking a lattice transcription
of the Chern-form. As a result, one finds (in four dimensions) an
action which depends on two holonomies. Furthermore, a crucial ingredient
is to consider these models for a general complex coupling parameter;
indeed, the interesting subdivision properties are present when the
associated scale parameter is a root of unity.
It should be emphasized that the Boltzmann weight itself is a group
valued object at these special points,
and this plays an important role in the analysis.

The novel features uncovered in these models should warrant further
investigation. Of most importance,
perhaps, is a detailed examination of the properties of the
Boltzmann weight under the remaining Alexander moves. The challenge
to come to a similar level of understanding for more complicated groups
is also well defined.
One would also like to see if a relation exists with the models
presented in \cite{DW,Alt}, and to explore the subdivision properties
of other correlation functions.

\end{document}